\documentclass[a4paper,fleqn]{cas-dc}
\usepackage[numbers]{natbib}
\usepackage{tabularx}

\makeatletter
\newcommand{\needspace}[1]{\begingroup\setlength{\dimen@}{#1}%
  \vskip\z@\@plus\dimen@ \penalty -100\vskip\z@\@plus -\dimen@
  \vskip\dimen@ \penalty 9999\vskip -\dimen@\endgroup}
\def\section{%
    \needspace{3\baselineskip}%
    \vskip 1.2ex%
    \hrule height 1.5pt%
    \@startsection{section}{1}{\z@}{-1pt}{1pt}{%
      \raggedright\normalfont\large\bfseries}%
}
\makeatother
\usepackage{multirow}
\usepackage{graphicx}
\graphicspath{ {./images/} }

\begin{document}
\let\WriteBookmarks\relax
\def\floatpagepagefraction{1}
\def\textpagefraction{.001}
\shorttitle{Detecting Privacy Policy Compliance with Data Protection Laws}
\shortauthors{A Qamar et~al.}
\title [mode = title]{Detecting Compliance of Privacy Policies with Data Protection Laws}

\author[1]{Ayesha Qamar}
\cormark[1]

\address[1]{National University of Computer and Emerging Sciences, Islamabad}

\author[1]{Tehreem Javed}
\cormark[1]

\author[1]{Mirza Omer Beg}

\nonumnote{\textit{Abbreviations}. NLP, natural language processing; GDPR, General Data Protection Regulation; PDPA, Personal Data Protection Act;  
  }

\cortext[cor1]{Corresponding authors: Tehreem Javed and Ayesha Qamar, Department of Computer Science, National University of Computer and Emerging Sciences, Islamabad\\
Email addresses: tehreem.javed202@gmail.com (T. Javed), ayeshaqamar55@gmail.com (A. Qamar), omer.beg@nu.edu.pk (M. Beg)}

\begin{abstract}
Privacy Policies are the legal documents that describe the practices that an organization or company has adopted in the handling of personal data of its users. But as policies are a legal document, they are often written in extensive legal jargon that is difficult to understand. Though work has been done on privacy policies but none that caters to the problem of verifying if a given privacy policy adheres to the data protection laws of a given country or state. We aim to bridge that gap by providing a framework that will analyse privacy policies in light of various data protection laws, such as the General Data Protection Regulation (GDPR). To achieve that, firstly we labelled both the privacy policies and laws. Then a correlation scheme is developed to map the contents of a privacy policy to the appropriate segments of law that a policy must conform to. Then we check the compliance of privacy policys' text with the corresponding text of the law using NLP techniques. By using such a tool, users would be better equipped to understand how their personal data is managed. For now, we have provided a mapping for the GDPR and PDPA, but other laws can easily be incorporated in the already built pipeline.
\end{abstract}


\maketitle

\section{Introduction}

In recent times, in the field of Natural Language Processing (or computer laws and policies?), a lot of work is being carried out to analyze\cite{Nokhbeh:2017}\cite{wilson:2016}\cite{Harkous:2018}\cite{Ramanath:2014}, understand\cite{Zimmeck:2019} and better represent\cite{Harkous:2018} privacy policies, none of the work targets to relate privacy policies with data protection laws. The analysis of privacy policies on their own is not enough. There needs to be a mechanism to relate those policies with laws. The policies dictate what an application or software is doing with the user's data but that information alone is not adequate to judge a policys' transparency and its usefulness \cite{cranor:2005}.

A possible solution is to create a system powered by machine learning to review the privacy policy and see if it is in accordance to the laws of the country (or countries) and identify any areas where a violation between them is detected. Using an automation tool, a user can have a deeper understanding of what is happening with their data in legal light.

The automation of checking compliance of privacy policies with laws can be of great value. It will arm users to understand policies with respect to laws without getting into the apprehension of legal jargon and details.

Privacy policies and data protection laws regulating these policies are both highly extensive and full of legal jargon. In fact, it is estimated that about 201 hours on average are needed by any average user just to read all the privacy policies encountered in a year\cite{mcdonald:2008}. As a result, consumers don't fully understand what they are signing up for\cite{rao:2016} and often do not know whether the policies that they are agreeing to are infringing on their legal rights.

Moreover, a company's legal department spends hours to review its privacy policy to see if it is compatible with a given country's laws. This is a rigorous process because each country has its own data protection laws and also because with the upsurge of Internet of Things there has been an escalation in the number and complexity of privacy policies themselves\cite{schaub:2017}. 

Hallinan et al\cite{Hallinan:2012} concluded through surveys that the European population at large remains skeptical now how their data is processed, any knowledge that the public has about data protection is superficial. In this technological era, users' understanding of how their data is processed is crucial for them to make informed decisions but users either don't have the basic understanding of their legal rights or not enough time to stay informed with the latest changes. This calls for a way to let users understand what they are signing up for without having to..


\section{Related work}

In 2016, Wilson et al\cite{wilson:2016} introduced a taxonomy for privacy policies, OPP-115, and made this corpus of 115 annotated policies publicly available. Since then, much work has been done to understand various aspects of privacy policies\cite{Harkous:2018}  \cite{Zimmeck:2019}. Sarne et al\cite{Sarne:2019} presented how using an unsupervised technique, Latent Dirichlet Allocation(LDA)\cite{Blei:2003}, can also provide a taxonomy for privacy policies that is much more fine-grained. LDA doesn't require the data to be pre-labelled. It works by randomly grouping words together into topics and then iteratively improving the grouping till convergence. They also showed that the taxonomy obtained had a substantial overlap with that of OOP-115. The research also provides insight into the topics that are being addressed in privacy policies these days. 

Apart from that, Hidden markov models\cite{Ramanath:2014} have been used previously to categorize privacy policies in an unsupervised way. The policies are segmented based on their section headings by crowdworkers. A Hidden Markov Model like approach is then used to align the segments such that an issue (addressed in the policy) corresponds to a hidden state. This correspondence is based on the bigrams in the segment of the policy and its distribution of words.

Tesfay et al\cite{Tesfay:2018} presented an approach to summarize long privacy policies using Machine Learning and then check against GDPR aspects as a criteria. Work has also been done to visually represent policies, for that Harkous et al\cite{Harkous:2018} developed a framework using Deep Learning techniques and the power of Convolutional Neural Networks to analyse policies on a finer level and developed a hierarchy to organise the information in privacy policies. They then presented the information in policies in a visual format and also provided a question answering interface where users' queries about a privacy policy are answered. 

Recently, Zimmeck et al\cite{Zimmeck:2019} compared the actual practice of a million apps with those stated in their privacy policies and flagged any discrepancies as compliance issues. 

Renaud K.\footnote{Data Protection Starter Kit. Retrieved June 4, 2020 from https://www.pdpc.gov.sg/-/media/files/pdpc/pdf-files/dp-starter-kit---171017.pdf} [23] analysis General Data Protection Regulation to find six requirements that a privacy policy compliant of GDPR must have. He provided a GDPR compliant privacy policy template for policy makers to use.

While work has been done to categorize, summarize and visualize privacy policies, none of the work has yet provided a universal method to check the privacy policies' compliance with the very laws that regulate them. 
\linebreak

\section{Methodology}

We propose a system which, given a privacy policy checks its compliance with a data protection law. For this, we first labelled privacy policies. Data protection laws were also segmented and labelled. Finally, we checked for the compliance of the resulting chunks of policies with those of the law. The details are mentioned in the following sections.

\subsection{Labelling Policies}
We have labelled policies based on the taxonomy provided by Wilson et al\cite{wilson:2016}. The taxonomy is based on a hierarchy of labelling and consists of 10 broad categories and 112 fine grained categories. The policies are segmented at paragraph level and each segment gets assigned multiple labels. The annotations were done by 3 graduate law students; there are three versions of the annotations. We have used the annotations in which there is a 0.75 overlap between the annotations, i.e., at least 2 of the 3 students have given the same label.

\begin{figure}[h!]
 \centering
  \includegraphics[width=3.2in]{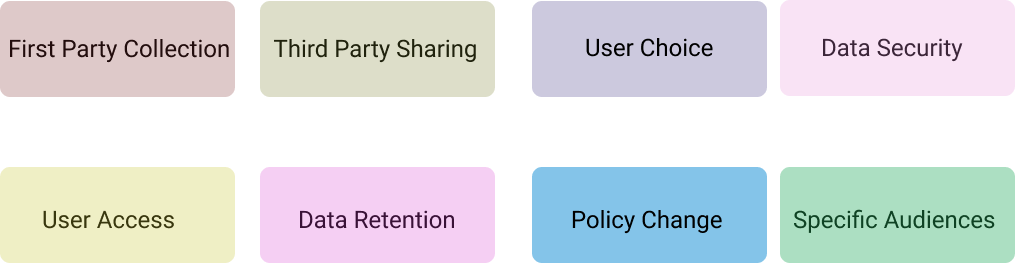}
  \caption{The 8 high level categories in the OPP-115 dataset. The other two categories not shown are \textit{other} and \textit{Do Not Track}. The latter category is not useful as it is no longer mentioned in policies.}
\end{figure}

To begin with, we extracted the 115 annotated policies from the OPP-115\cite{wilson:2016} dataset, and only relevant information like the text segment of policies themselves and the assigned labels were kept.  Then to cater for these multi class labels, we made binary models for classification for the 10 broad categories. Thus, for training the classifiers the dataset was divided into ten subsets where each set corresponded to one category and had binary labels (0 if the text segment did not belong to the category and 1 otherwise).

Then for the classification, we used Towards Automatic Classification of Policy Text\cite{Liu:2017} as a starting point and trained a Logistic Regression model and a Support Vector Machine model for classification. In addition, we also used a fine tuned version of the BERT\cite{Devlin:2018} model. We took the pre-trained BERT model for classification and fine tuned it using a low learning rate for each policy category. We trained classifiers for all the ten datasets and calculated their F1 scores. 

We tested the three models for each category on a held out test set and calculated their F1 scores. The BERT Classifier gave better results than others, so we saved the trained model to use for privacy policy categorization at run time in the final product.

\subsection{Labelling Laws}
We have worked with two laws i.e., the General Data Protection Regulation-GDPR and the Personal Data Protection Act-PDPA. The first step to labelling the laws is to segment them.

\subsubsection{General Data Protection Regulation}
For the GDPR, we followed the natural hierarchy in which it is written and segmented it according to the Articles, with one segment consisting of all the subpoints of an Article. By following this segmentation scheme we were left with 371 segments with an average word count of 75.11 words per segment. After that, we removed stopwords, punctuations and lemmatized the words. 

\begin{figure}[h!]
 \centering
  \includegraphics[width=3.2 in]{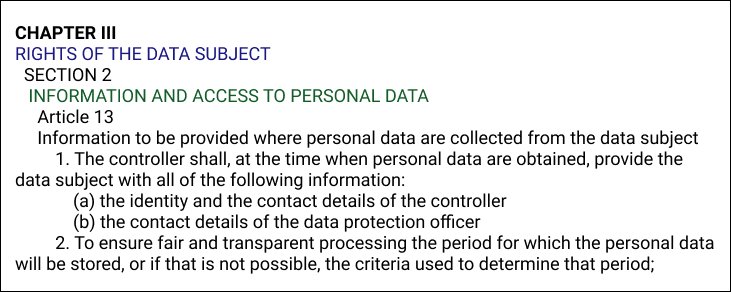}
  \caption{The structure of the GDPR. The hierarchy consists of Chapters, Sections, Articles and then points in those Articles.}
\end{figure}

We decided to use topic modelling for grouping together similar segments and thus creating a taxonomy of the law. We used Latent Dirichlet Allocation (LDA)\cite{Blei:2003} to achieve that. The decision to use LDA was based on the promising results achieved by\cite{Sarne:2019} to label privacy policies. LDA works by assuming that topics in a document and words in a topic follow some specific distribution. Since it's an unsupervised technique, we only need to provide the number of topics, \textit{k}, the document has. Since it's a hyperparameter, we experimented with several values of \textit{k} and found that setting it to 10 gave the most optimal results in our case.

\begin{figure}[h!]
 \centering
  \includegraphics[width=3.2 in]{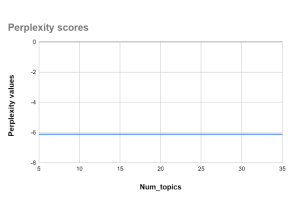}
  \caption{The perplexity score plotted against multiple values of the number of topics.}
\end{figure}

Perplexity scores(the lower, the better) did not give any insightful information to decide the value of \textit{k}. Therefore, we used the coherence score(a measure to evaluate topic models) as the deciding factor instead. The best coherence value was achieved when k was set to 5. But such a coarse labelling would not have served our purpose, since we know that the GDPR contains at least 10 different topics as those are the number of different chapters, so we went with the next favourable value of 15. 

\begin{figure}[h!]
 \centering
  \includegraphics[width=3.2 in]{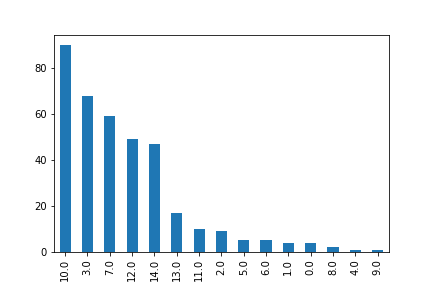}
  \caption{The number of segments belonging to each topic is depicted when \textit{k} is set to 15. As shown, some of the topics only have one or two segments assigned to them. }
\end{figure}

But setting the number of topics to 15 gave rise to a few topics containing only one or two segments only and merging them seemed to be a sensible option. So in the end we decided to keep the number of topics to 10.

\begin{figure}[h!]
 \centering
  \includegraphics[width=3.2 in]{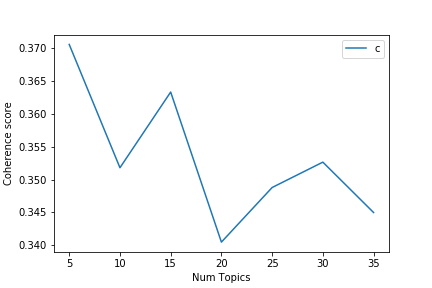}
  \caption{The coherence scores plotted against the number of topics \textit{k}.}
\end{figure}

\begin{figure}[h!]
 \centering
  \includegraphics[width=3.2 in]{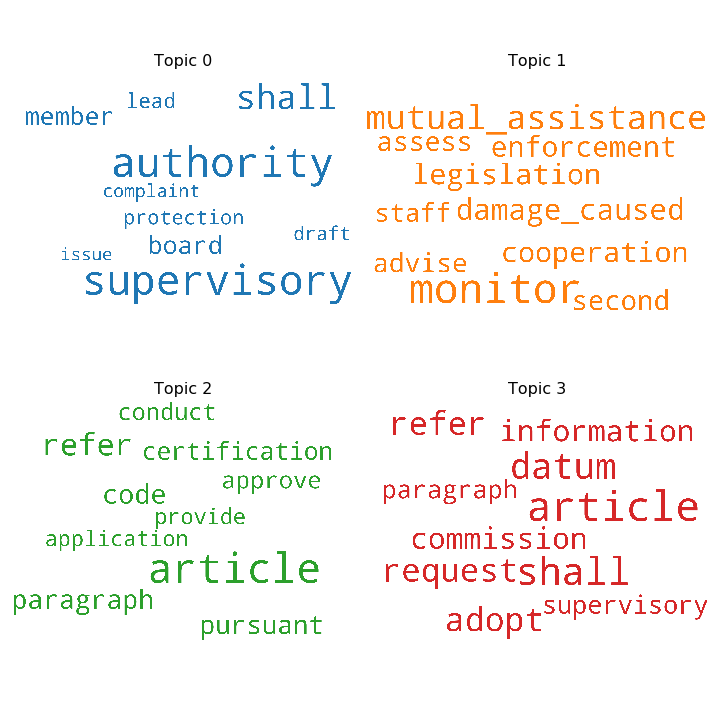}
  \caption{A visual representation of the most frequent words of some of the topics. }
\end{figure}

Figure 6 shows the most occurring words in four of the topics. Most of the words are non-overlapping i.e., do not occur in multiple topics and hence show that the labelling is efficient.

\subsubsection{Personal Data Protection Act}

The PDPA has two main provisions: 
\begin{description}
  \item[$\bullet$ Data Protection (DP) provisions.] These provisions are directly concerned with the handling and collection of users' personal data..
  \item[$\bullet$ Do Not Call (DNC) provisions.] Do Not Call Registry is not applicable to privacy policies as that part is only concerned with how to handle the telephone numbers of Singaporean users but does not in particular detail how the phone number should be collected. Including video or voice calls or text messages. But since these requirements aren't directly linked with privacy policies, we skipped those divisions.

\end{description}

For the PDPA, we did manual annotation. As the PDPA is a relatively shorter law, we did not feel the need to label it through any unsupervised method to obtain the segment categorizations. Upon manual reading only PART II to PART V of the law were relevant to personal data and we extracted appropriate text from these parts. 

Titles of the parts, from where law text was extracted, are mentioned below \footnote{PERSONAL DATA PROTECTION ACT 2012.  Retrieved June 10, 2020 from https://sso.agc.gov.sg/Act/PDPA2012}[22]:
\begin{description}
  \item[$\bullet$ PART II]: \textit{PERSONAL DATA PROTECTION COMMISSION AND ADMINISTRATION }
  \item[$\bullet$ PART III]: \textit{GENERAL RULES WITH RESPECT TO PROTECTION OF PERSONAL DATA }
  \item[$\bullet$ PART IV]: \textit{COLLECTION, USE AND DISCLOSURE OF PERSONAL DATA}
   \item[$\bullet$ PART IV]: \textit{PART V: ACCESS TO AND CORRECTION OF PERSONAL DATA}
\end{description}

\subsection{Extracting Relevant Law Segments}

\subsubsection{General Data Protection Regulation}
Leveraging the work done by Karen et al\cite{Renaud:2018}, where they provide a template and lay out the requirements that policies must follow in order to be GDPR compliant. The GDPR requirements that customers must be informed about are:
\begin{description}
  \item[$\bullet$ GDPR1]: \textit{What Data will be Collected and Why\footnote{We merged two categories from the paper into one.}}
  \item[$\bullet$ GDPR2]: \textit{How Data Will Be Processed}
  \item[$\bullet$ GDPR3]: \textit{How Long Data Will Be Retained}
  \item[$\bullet$ GDPR4]: \textit{Who Can Be Contacted to Have Data Removed or Produced}
\end{description}


Next we were left with the task of manually extracting all the text from the GDPR that pertained to the specific category of our interest. Because the law was already categorized using LDA, this step became easier. Only some portion of the law was useful for our purpose i.e., the articles related to personal data processing and not the chapters about how the law itself should be implemented or where it is applicable such as the Chapter X: DELEGATED ACTS AND IMPLEMENTING ACTS.

For example, the Article 14 of GDPR on Information to be provided where personal data have not been obtained from the data subject\footnote{ GDPR Article 14(1)(d,e):“Where personal data have not been obtained from the data subject, the controller shall provide the data subject with the following information: the categories of personal data concerned the recipients or categories of recipients of the personal data where applicable”} [21]??, this was categorized as a GDPR segment belonging to “What Data will be Collected and Why” and so will be mapped to the First Party and Third Party category of privacy policy. In total, ten such law segments were made and given one of the four above mentioned requirements.

\subsubsection{Personal Data Protection Act}
According to the Personal Data Protection Commission (PDPC), there are three broader categories and then further subcategories of the obligations set by PDPA regarding protection of personal data\footnote{Data Protection Starter Kit. Retrieved June 4, 2020 from https://www.pdpc.gov.sg/-/media/files/pdpc/pdf-files/dp-starter-kit---171017.pdf} [23]. We extracted only those subcategories that applied to privacy policies. The categories with a brief description are stated below:

\begin{itemize}
   \item \textbf{Collection, use and disclosure of personal data}
 
   \begin{itemize}
     \item \textbf{\textit{Consent}}
     An organization should first ask customers to give consent to collect, use or disclose their personal data. Users should also have the ability to withdraw consent.
    \item \textbf{\textit{Purpose and Notification}}
    Consent should only be taken for data that is essential to provide a given service to users. Users' data can only be obtained or disclosed for the purposes for which the user was informed about. Users should also be made aware of the reasons for data collection.
   \end{itemize}
 
   \item \textbf{Accountability to individuals}
    \begin{itemize}
     \item \textbf{\textit{Access and Correction}}
    If customers request, they should be provided with their collected personal data along with the ways in which the data was collected and used in the time frame of a year. Users can also request to get their data updated to fix any errors.
   \end{itemize}

   \item \textbf{Care of personal data}
       \begin{itemize}
     \item \textbf{\textit{Retention}}
    Data should be deleted once the purpose it was collected for has been fulfilled. Keeping data longer than needed for business reasons is prohibited.

   \end{itemize}
 
\end{itemize}

  


\subsection{Correlating Policy and Law Segments}
When a new policy is entered to check for compliance, first it is segmented and then each segment gets labelled one or more of the 10 labels. At this stage, we have categorized policy segments(entered at runtime) and have the already labelled law segments. 
Next, we need to relate each of the policy segments with one or more of the law segments\\

\textbf{General Data Protection Regulation}

For GDPR, we take the labelled segments of policy and map the five privacy policy categories to the four GDPR requirement categories.

\begin{figure}[h!]
 \centering
  \includegraphics[width=3.2 in]{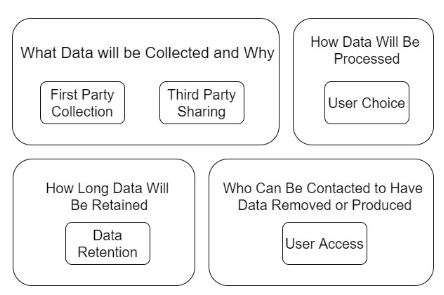}
  \caption{The GDPR law segments categories represented by the outer box and the policy categories linked with each.}
\end{figure}

\textbf{Personal Data Protection Act}

The policy categories are mapped to PDPA based on the guideline provided by the PDPC
\begin{figure}[h!]
 \centering
  \includegraphics[width=3.2 in]{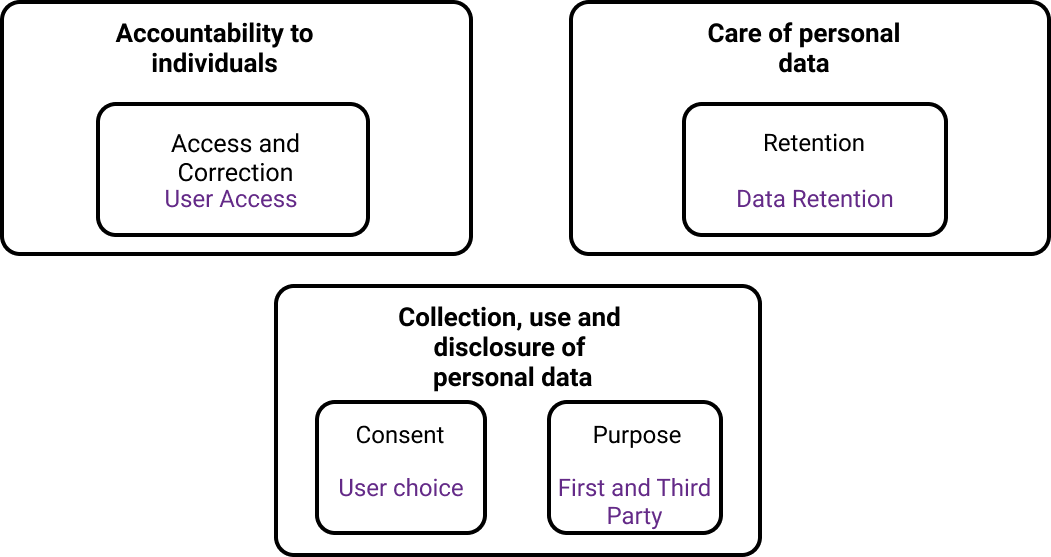}
  \caption{The PDPA law categories represented by the outer box and the policy categories written in purple.}
\end{figure}

\subsection{Finding Similarity}
After allocating categories to segments of laws and policies, we find similarity between segments of the laws and policies which fall under the same category. This similarity is used as a measure to decide if the policy is in compliance with the law. We used BERT \cite{Devlin:2018}[17] word embeddings and Universal Sentence encoding[18] to find the similarity. 

Word embeddings such as word2vec and Glove have been useful in improving accuracy across NLP tasks. BERT word embeddings improve upon these methods because it is the first unsupervised, deeply bidirectional system for pre-training NLP. Context-free models such as word2vec or GloVe generate a single "word embedding" representation for each word in the vocabulary, so bank would have the same representation in bank deposit and river bank. We use pre trained BERT uncased model to get sentence embeddings by combining word embeddings through mean across layers of words. These embeddings are then used to find similarity between pairs of policy and law segment using cosine similarity and euclidean distance.

By using Universal Sentence Encoding\cite{Cer:2018}, we obtained sentence embeddings and then used cosine similarity. There are two model architectures present, one uses transformer architecture and gets higher quality embeddings while requiring greater resources and computing power and the second one uses less resources but at the cost of slightly lesser accuracy. We went with the latter one to utilize resources optimally. The architecture we used is the Deep Averaging Network (DAN), first word embeddings along with bi-grams are averaged and then used as input to feedforward deep neural network (DNN) to get sentence embeddings of 512 dimensions.

\section{Compliance Score}

Using the similarity score between the policy segments and the law segments which are related to each other, as the starting point we calculate the compliance score using the formula shown below:

\begin{align*}
  Compliance &= \frac{Max - Score}{Max - Min}\\
\end{align*}

As we don't have a labelled dataset for compliance score between law and policy segment, we created a small set to find the required compliance thresholds(max and min) .To decide on where to set the threshold for compliance and non-compliance from the cosine similarity score obtained from Universal Sentence Encodings of policy and law segments, we created a dummy dataset. The dataset consists of a law segment, for both PDPA and GDPR, and policy segment along with a score from 0 to 5; 1 being the least compliant, 5 being completely compliant and 0 showing absolute irrelevance between texts. Then the problem simply reduced to identifying the correct value of thresholds to turn the similarity score into compliance score. 

For the GDPR, the threshold was found to be max 0.6 and minimum 0.25, that is, when a policy segment was in complete compliance of a law segment the similarity score was 0.6 and when it had zero compliance the score was 0.25. Using these thresholds, we find the compliance score for the four GDPR requirements; what data will be collected and why, how data will be processed, how long it will be retained and who can be contacted to have data removed or produced.

For the PDPA, the thresholds that gave the optimal results were max .09 for total compliance and min .21 for ½ compliance, with the compliance decrementing as the score increased towards .50.
\linebreak

\section{Experimental Evaluation}

\begin{description}
  \item[$\bullet$ Labelling Policies:] To evaluate the labelling of privacy policies we used a held out dataset and checked the accuracy of our models(SVM, LR, BERT) on that data. The complete results can be seen in figure 9. BERT gave a better F1 score for most categories.  
  \item[$\bullet$ Labelling Laws:] For laws, we are going to have an expert verify the labelling and annotation since there is no labelled dataset of data protection laws available.
  \item [$\bullet$ Finding Similarity:] Due to the unavailability of policy and law compliance dataset, we evaluate our similarity model by using it on the semantic textual similarity development dataset. The STS dataset comprises of sentence pairs from news, captions, and forums genre. These sentence pairs are labelled for similarity on a scale of 0 to 5 where 5 means complete similarity and 0 means no similarity at all. The Pearson Correlation obtained by using BERT embedding and taking mean of all word vectors and sum of all vectors as well as the correlation obtained by using Universal Sentence encoding is shown in figure 8.
   \item [$\bullet$ Test Case:]
   We tested our system by using the nestle privacy policy. This policy contains a clause about data retention which states “Nestlé will only retain your personal data for as long as it is necessary for the stated purpose, taking into account also our need to answer queries or resolve problems, provide improved and new services, and comply with legal requirements under applicable laws.This means that we may retain your personal data for a reasonable period after your last interaction with us. When the personal data that we collect is no longer required in this way, we destroy or delete it in a secure manner.”
   
We first run the privacy policy against GDPR as it is and the system gives a 99.7 percent data retention compliance score as it should. Then we replace this section with “Nestle will store the data for as long as we want”. When this altered policy is run against GDPR, the compliance report gives a score of around zero percent.

\end{description}

\begin{table}[ht]
\begin{center}
    \caption{The pearson correlation obtained on Universal Sentence Encoding outperforms BERT models.}
    \begin{tabular}{  l  l  c  p{5cm}}
    \hline
    \bf Model &\bf  Pearson Correlation \\[5pt] \hline    \hline
    BERT with Cosine Similarity & 0.55\\[5pt]
    BERT with euclidean distance & 0.57 \\[5pt]
    Universal Sentence Encoder & 0.76  \\[5pt]
    \hline
    \end{tabular}
\end{center}
\end{table}


\begin{table}[h]
\caption{The F1 score of the Logistic Regression, Support Vector Machine and BERT across all the categories.}
\begin{tabular}{ c c c c }
\hline
\bf Categories & LR & SVM & BERT \\[2pt]
\hline \hline
 First Party Collection/Use & 0.79 & 0.81 & 0.85 \\[1pt]
 Third Party Sharing/Collection & 0.77 & 0.78 & 0.87 \\
 User Choice/Control & 0.68 & 0.70 & 0.73 \\
 User Access, Edit and Deletion & 0.81 & 0.82 & 0.66 \\
 Data Retention & 0.43 & 0.40 & 0.62 \\
 Data Security & 0.73 & 0.77 & 0.77 \\
\hline
\end{tabular}

\end{table}

\section{Conclusion and Further Work}
Our work proves that automated compliance check with regard to legal documents gives plausible results. This opens the possibility of using such techniques to find legal compliance in contracts etc. Further work can be done in this area by adding further data protection laws such as Canada's PIPEDA and US' Privacy Shield. Another improvement that can be done is to try more complex architectures and models to correlate laws and policies.

\nocite{*}
\bibliographystyle{abbrvnat}

\begin{thebibliography}{}
\bibitem{mcdonald:2008}
A. M. McDonald and L. F. Cranor, “The cost of reading privacy policies,” ISJLP, vol. 4, p. 543, 2008. 
\bibitem{rao:2016}
A. Rao, F. Schaub, N. Sadeh, A. Acquisti, and R. Kang, “Expecting the unexpected: Understanding mismatched privacy expectations online,” in Twelfth Symposium on Usable Privacy and Security (SOUPS 2016). Denver, CO: USENIX Association, 2016, pp. 77–96. 
\bibitem{schaub:2017}
F. Schaub, R. Balebako, and L. F. Cranor, “Designing effective privacy notices and controls,” IEEE Internet Computing, vol. 21, no. 3, pp. 70–77, 2017. 
\bibitem{cranor:2005}
Cranor, Lorrie Faith. "Giving notice: why privacy policies and security breach notifications aren't enough." IEEE Communications Magazine 43.8 (2005): 18-19.
\bibitem{wilson:2016}
Wilson, Shomir, et al. "The creation and analysis of a website privacy policy corpus." Proceedings of the 54th Annual Meeting of the Association for Computational Linguistics (Volume 1: Long Papers). 2016.
\bibitem{sarne:2019}
Sarne, D., Schler, J., Singer, A., Sela, A. and Bar Siman Tov, I., 2019, May. Unsupervised Topic Extraction from Privacy Policies. In Companion Proceedings of The 2019 World Wide Web Conference (pp. 563-568). ACM
\bibitem{Angeli:2015}
Angeli, Gabor, Melvin Jose Johnson Premkumar, and Christopher D. Manning. "Leveraging linguistic structure for open domain information extraction." Proceedings of the 53rd Annual Meeting of the Association for Computational Linguistics and the 7th International Joint Conference on Natural Language Processing (Volume 1: Long Papers). 2015.
\bibitem{Linden:2018}
Linden, Thomas, et al. "The privacy policy landscape after the GDPR." arXiv preprint arXiv:1809.08396 (2018).
\bibitem{Jensen:2004}
Jensen, Carlos, and Colin Potts. "Privacy policies as decision-making tools: an evaluation of online privacy notices." Proceedings of the SIGCHI conference on Human Factors in Computing Systems. ACM, 2004.
\bibitem{}
M. Hochhauser (2001). Lost in the fine print: Readability of financial privacy notices. Retrieved September 30, 2019 from http://www.privacyrights.org/ar/GLB-Reading.htm. 
\bibitem{Antón:2002}
Antón, Annie I., Julia Brande Earp, and Angela Reese. "Analyzing website privacy requirements using a privacy goal taxonomy." Proceedings IEEE Joint International Conference on Requirements Engineering. IEEE, 2002.
\bibitem{Harkous:2018}
Harkous, Hamza, et al. "Polisis: Automated analysis and presentation of privacy policies using deep learning." 27th {USENIX} Security Symposium ({USENIX} Security 18). 2018.
\bibitem{Sarne:2019}
Sarne, D., Schler, J., Singer, A., Sela, A. and Bar Siman Tov, I., 2019, May. Unsupervised Topic Extraction from Privacy Policies. In Companion Proceedings of The 2019 World Wide Web Conference (pp. 563-568). ACM
\bibitem{Ramanath:2014}
Ramanath, Rohan, et al. "Unsupervised alignment of privacy policies using hidden markov models." Proceedings of the 52nd Annual Meeting of the Association for Computational Linguistics (Volume 2: Short Papers). 2014.

\bibitem{Hallinan:2012}
Hallinan, Dara, Michael Friedewald, and Paul McCarthy. "Citizens' perceptions of data protection and privacy in Europe." Computer law \& security review 28.3 (2012): 263-272.

\bibitem{Nokhbeh:2017}
Nokhbeh Zaeem, R., \& Barber, K. S. (2017). A study of web privacy policies across industries. Journal of Information Privacy and Security, 13(4), 169-185.

\bibitem[Alvi et~al.(2017)Alvi, Sahar, Bangash, and Beg]{alvi2017ensights}
H.~M. Alvi, H.~Sahar, A.~A. Bangash, and M.~O. Beg.
\newblock Ensights: A tool for energy aware software development.
\newblock In \emph{2017 13th International Conference on Emerging Technologies
  (ICET)}, pages 1--6. IEEE, 2017.

\bibitem[Anwar and Baig(2020)]{anwar2020tac}
T.~Anwar and O.~Baig.
\newblock Tac at semeval-2020 task 12: Ensembling approach for multilingual
  offensive language identification in social media.
\newblock In \emph{Proceedings of the Fourteenth Workshop on Semantic
  Evaluation}, pages 2177--2182, 2020.

\bibitem[Arshad et~al.(2019)Arshad, Bashir, Majeed, Shahzad, and
  Beg]{arshad2019corpus}
M.~U. Arshad, M.~F. Bashir, A.~Majeed, W.~Shahzad, and M.~O. Beg.
\newblock Corpus for emotion detection on roman urdu.
\newblock In \emph{2019 22nd International Multitopic Conference (INMIC)},
  pages 1--6. IEEE, 2019.

\bibitem[Asad et~al.(2020)Asad, Asim, Javed, Beg, Mujtaba, and
  Abbas]{asad2020deepdetect}
M.~Asad, M.~Asim, T.~Javed, M.~O. Beg, H.~Mujtaba, and S.~Abbas.
\newblock Deepdetect: detection of distributed denial of service attacks using
  deep learning.
\newblock \emph{The Computer Journal}, 63\penalty0 (7):\penalty0 983--994,
  2020.

\bibitem[Awan and Beg(2021)]{awan2021top}
M.~N. Awan and M.~O. Beg.
\newblock Top-rank: a topicalpostionrank for extraction and classification of
  keyphrases in text.
\newblock \emph{Computer Speech \& Language}, 65:\penalty0 101--116, 2021.

\bibitem[Bangash et~al.(2017)Bangash, Sahar, and Beg]{bangash2017methodology}
A.~A. Bangash, H.~Sahar, and M.~O. Beg.
\newblock A methodology for relating software structure with energy
  consumption.
\newblock In \emph{2017 IEEE 17th International Working Conference on Source
  Code Analysis and Manipulation (SCAM)}, pages 111--120. IEEE, 2017.

\bibitem[Beg(2008)]{beg2008critical}
M.~Beg.
\newblock Critical path heuristic for automatic parallelization.
\newblock \emph{University of Waterloo, David R. Cheriton School of Computer
  Science, Technical Report CS-2008-16}, 2008.

\bibitem[Beg and Beek(2013)]{beg2013constraint}
M.~Beg and P.~v. Beek.
\newblock A constraint programming approach for integrated spatial and temporal
  scheduling for clustered architectures.
\newblock \emph{ACM Transactions on Embedded Computing Systems (TECS)},
  13\penalty0 (1):\penalty0 1--23, 2013.

\bibitem[Beg and Dahlin(2001)]{beg2001memory}
M.~Beg and M.~Dahlin.
\newblock A memory accounting interface for the java programming language.
\newblock \emph{Technical Report CS-TR-01--40, University of Texas at Austin},
  2001.

\bibitem[Beg and Van~Beek(2010)]{beg2010graph}
M.~Beg and P.~Van~Beek.
\newblock A graph theoretic approach to cache-conscious placement of data for
  direct mapped caches.
\newblock In \emph{Proceedings of the 2010 international symposium on Memory
  management}, pages 113--120, 2010.

\bibitem[Beg and Van~Beek(2011)]{beg2011constraint}
M.~Beg and P.~Van~Beek.
\newblock A constraint programming approach for instruction assignment.
\newblock In \emph{2011 15th Workshop on Interaction between Compilers and
  Computer Architectures}, pages 25--34. IEEE, 2011.

\bibitem[Beg(2013)]{beg2013combinatorial}
M.~O. Beg.
\newblock Combinatorial problems in compiler optimization.
\newblock 2013.

\bibitem[Beg et~al.(2019)Beg, Awan, and Ali]{beg2019algorithmic}
M.~O. Beg, M.~N. Awan, and S.~S. Ali.
\newblock Algorithmic machine learning for prediction of stock prices.
\newblock In \emph{FinTech as a Disruptive Technology for Financial
  Institutions}, pages 142--169. IGI Global, 2019.

\bibitem[Dilawar et~al.(2018)Dilawar, Majeed, Beg, Ejaz, Muhammad, Mehmood, and
  Nam]{dilawar2018understanding}
N.~Dilawar, H.~Majeed, M.~O. Beg, N.~Ejaz, K.~Muhammad, I.~Mehmood, and Y.~Nam.
\newblock Understanding citizen issues through reviews: A step towards data
  informed planning in smart cities.
\newblock \emph{Applied Sciences}, 8\penalty0 (9):\penalty0 1589, 2018.

\bibitem[Farooq et~al.(2019{\natexlab{a}})Farooq, Beg,
  et~al.]{farooq2019bigdata}
M.~U. Farooq, M.~O. Beg, et~al.
\newblock Bigdata analysis of stack overflow for energy consumption of android
  framework.
\newblock In \emph{2019 International Conference on Innovative Computing
  (ICIC)}, pages 1--9. IEEE, 2019{\natexlab{a}}.

\bibitem[Farooq et~al.(2019{\natexlab{b}})Farooq, Khan, and
  Beg]{farooq2019melta}
M.~U. Farooq, S.~U.~R. Khan, and M.~O. Beg.
\newblock Melta: A method level energy estimation technique for android
  development.
\newblock In \emph{2019 International Conference on Innovative Computing
  (ICIC)}, pages 1--10. IEEE, 2019{\natexlab{b}}.

\bibitem[Javed et~al.(2020{\natexlab{a}})Javed, Beg, Asim, Baker, and
  Al-Bayatti]{javed2020alphalogger}
A.~R. Javed, M.~O. Beg, M.~Asim, T.~Baker, and A.~H. Al-Bayatti.
\newblock Alphalogger: Detecting motion-based side-channel attack using
  smartphone keystrokes.
\newblock \emph{Journal of Ambient Intelligence and Humanized Computing}, pages
  1--14, 2020{\natexlab{a}}.

\bibitem[Javed et~al.(2020{\natexlab{b}})Javed, Sarwar, Beg, Asim, Baker, and
  Tawfik]{javed2020collaborative}
A.~R. Javed, M.~U. Sarwar, M.~O. Beg, M.~Asim, T.~Baker, and H.~Tawfik.
\newblock A collaborative healthcare framework for shared healthcare plan with
  ambient intelligence.
\newblock \emph{Human-centric Computing and Information Sciences}, 10\penalty0
  (1):\penalty0 1--21, 2020{\natexlab{b}}.

\bibitem[Javed et~al.(2019)Javed, Beg, Mujtaba, Majeed, and
  Asim]{javed2019fairness}
H.~T. Javed, M.~O. Beg, H.~Mujtaba, H.~Majeed, and M.~Asim.
\newblock Fairness in real-time energy pricing for smart grid using
  unsupervised learning.
\newblock \emph{The Computer Journal}, 62\penalty0 (3):\penalty0 414--429,
  2019.

\bibitem[Karsten et~al.(2007)Karsten, Keshav, Prasad, and
  Beg]{karsten2007axiomatic}
M.~Karsten, S.~Keshav, S.~Prasad, and M.~Beg.
\newblock An axiomatic basis for communication.
\newblock \emph{ACM SIGCOMM Computer Communication Review}, 37\penalty0
  (4):\penalty0 217--228, 2007.

\bibitem[Khawaja et~al.(2018)Khawaja, Beg, and Qamar]{khawaja2018domain}
H.~S. Khawaja, M.~O. Beg, and S.~Qamar.
\newblock Domain specific emotion lexicon expansion.
\newblock In \emph{2018 14th International Conference on Emerging Technologies
  (ICET)}, pages 1--5. IEEE, 2018.

\bibitem[Majeed et~al.(2020)Majeed, Mujtaba, and Beg]{majeed2020emotion}
A.~Majeed, H.~Mujtaba, and M.~O. Beg.
\newblock Emotion detection in roman urdu text using machine learning.
\newblock In \emph{Proceedings of the 35th IEEE/ACM International Conference on
  Automated Software Engineering Workshops}, pages 125--130, 2020.

\bibitem[Naeem et~al.(2020{\natexlab{a}})Naeem, Khan, Beg, and
  Mujtaba]{naeem2020deep}
B.~Naeem, A.~Khan, M.~O. Beg, and H.~Mujtaba.
\newblock A deep learning framework for clickbait detection on social area
  network using natural language cues.
\newblock \emph{Journal of Computational Social Science}, pages 1--13,
  2020{\natexlab{a}}.

\bibitem[Naeem et~al.(2020{\natexlab{b}})Naeem, Iqbal, Saqib, Saad, Raza, Ali,
  Akhtar, Beg, Shahzad, and Arshad]{naeem2020subspace}
S.~Naeem, M.~Iqbal, M.~Saqib, M.~Saad, M.~S. Raza, Z.~Ali, N.~Akhtar, M.~O.
  Beg, W.~Shahzad, and M.~U. Arshad.
\newblock Subspace gaussian mixture model for continuous urdu speech
  recognition using kaldi.
\newblock In \emph{2020 14th International Conference on Open Source Systems
  and Technologies (ICOSST)}, pages 1--7. IEEE, 2020{\natexlab{b}}.

\bibitem[Qamar et~al.(2021)Qamar, Mujtaba, Majeed, and Beg]{qamarrelationship}
S.~Qamar, H.~Mujtaba, H.~Majeed, and M.~O. Beg.
\newblock Relationship identification between conversational agents using
  emotion analysis.
\newblock \emph{Cognitive Computation}, pages 1--15, 2021.

\bibitem[Sahar et~al.(2019)Sahar, Bangash, and Beg]{sahar2019towards}
H.~Sahar, A.~A. Bangash, and M.~O. Beg.
\newblock Towards energy aware object-oriented development of android
  applications.
\newblock \emph{Sustainable Computing: Informatics and Systems}, 21:\penalty0
  28--46, 2019.

\bibitem[Tariq et~al.(2019)Tariq, Majeed, Beg, Khan, and
  Derhab]{tariq2019accurate}
M.~Tariq, H.~Majeed, M.~O. Beg, F.~A. Khan, and A.~Derhab.
\newblock Accurate detection of sitting posture activities in a secure iot
  based assisted living environment.
\newblock \emph{Future Generation Computer Systems}, 92:\penalty0 745--757,
  2019.

\bibitem[Uzair et~al.(2019)Uzair, Beg, Mujtaba, and Majeed]{uzair2019weec}
A.~Uzair, M.~O. Beg, H.~Mujtaba, and H.~Majeed.
\newblock Weec: Web energy efficient computing: A machine learning approach.
\newblock \emph{Sustainable Computing: Informatics and Systems}, 22:\penalty0
  230--243, 2019.

\bibitem[Zafar et~al.(2018)Zafar, Mujtaba, Beg, and Ali]{zafar2018deceptive}
A.~Zafar, H.~Mujtaba, M.~O. Beg, and S.~Ali.
\newblock Deceptive level generator.
\newblock In \emph{AIIDE Workshops}, 2018.

\bibitem[Zafar et~al.(2019{\natexlab{a}})Zafar, Mujtaba, Ashiq, and
  Beg]{zafar2019constructive}
A.~Zafar, H.~Mujtaba, S.~Ashiq, and M.~O. Beg.
\newblock A constructive approach for general video game level generation.
\newblock In \emph{2019 11th Computer Science and Electronic Engineering
  (CEEC)}, pages 102--107. IEEE, 2019{\natexlab{a}}.

\bibitem[Zafar et~al.(2019{\natexlab{b}})Zafar, Mujtaba, Baig, and
  Beg]{zafar2019using}
A.~Zafar, H.~Mujtaba, M.~T. Baig, and M.~O. Beg.
\newblock Using patterns as objectives for general video game level generation.
\newblock \emph{ICGA Journal}, 41\penalty0 (2):\penalty0 66--77,
  2019{\natexlab{b}}.

\bibitem[Zafar et~al.(2020)Zafar, Mujtaba, and Beg]{zafar2020search}
A.~Zafar, H.~Mujtaba, and M.~O. Beg.
\newblock Search-based procedural content generation for gvg-lg.
\newblock \emph{Applied Soft Computing}, 86:\penalty0 105909, 2020.

\bibitem{He:2015}
He, Hua, Kevin Gimpel, and Jimmy Lin. "Multi-perspective sentence similarity modeling with convolutional neural networks." Proceedings of the 2015 Conference on Empirical Methods in Natural Language Processing. 2015.
\bibitem{Zimmeck:2019}
Zimmeck, Sebastian, et al. "MAPS: Scaling privacy compliance analysis to a million apps." Proceedings on Privacy Enhancing Technologies 2019.3 (2019): 66-86.
\bibitem{Devlin:2018}
Devlin, J., Chang, M. W., Lee, K., \& Toutanova, K. (2018). Bert: Pre-training of deep bidirectional transformers for language understanding. arXiv preprint arXiv:1810.04805.
\bibitem{Cer:2018}
Cer, D., Yang, Y., Kong, S. Y., Hua, N., Limtiaco, N., John, R. S., ... \& Sung, Y. H. (2018). Universal sentence encoder. arXiv preprint arXiv:1803.11175.
\bibitem{Liu:2017}
Liu, F., Wilson, S., Story, P., Zimmeck, S. and Sadeh, N., 2017. Towards Automatic Classification of Privacy Policy Text.
\bibitem{Tesfay:2018}
Tesfay, W. B., Hofmann, P., Nakamura, T., Kiyomoto, S., \& Serna, J. (2018, April). I Read but Don't Agree: Privacy Policy Benchmarking using Machine Learning and the EU GDPR. In Companion Proceedings of the The Web Conference 2018 (pp. 163-166). International World Wide Web Conferences Steering Committee.
\bibitem{Blei:2003}
Blei, D. M., Ng, A. Y., \& Jordan, M. I. (2003). Latent dirichlet allocation. Journal of machine Learning research, 3(Jan), 993-1022.

\bibitem{}
PERSONAL DATA PROTECTION ACT 2012.  Retrieved June 10, 2020 from https://sso.agc.gov.sg/Act/PDPA2012

\bibitem{Renaud:2018}
Renaud, Karen, and Lynsay A. Shepherd. "How to make privacy policies both GDPR-compliant and usable." 2018 International Conference On Cyber Situational Awareness, Data Analytics And Assessment (Cyber SA). IEEE, 2018.

\bibitem{}
Data Protection Starter Kit. Retrieved June 4, 2020 from https://www.pdpc.gov.sg/-/media/files/pdpc/pdf-files/dp-starter-kit---171017.pdf


\end{thebibliography}



\end{document}